\documentclass[aps,prl, twocolumn,superscriptaddress,preprintnumbers,floatfix,nofootinbib,notitlepage]{revtex4-1}

\usepackage{graphicx}
\usepackage{amsmath,amsfonts,amssymb}
\usepackage[colorlinks,urlcolor=black,citecolor=blue,linkcolor=black]{hyperref}
\usepackage{color}
\usepackage{verbatim}
\usepackage{enumitem}

\def\lsim{\mathrel{\raise.3ex\hbox{$<$\kern-.75em\lower1ex\hbox{$\sim$}}}}
\def\gsim{\mathrel{\raise.3ex\hbox{$>$\kern-.75em\lower1ex\hbox{$\sim$}}}}

\newcommand{\be}{\begin{equation}}
\newcommand{\ee}{\end{equation}}
\newcommand{\bea}{\begin{equation}\begin{aligned}}
\newcommand{\eea}{\end{aligned}\end{equation}}
\newcommand{\td}{{\rm d}}

\newcommand{\Msun}{M_{\odot}}

\newcommand{\Mpc}{{\rm Mpc}}

\newcommand{\GeV}{{\rm GeV}}
\newcommand{\Mpl}{M_{\rm P}}

\setlist[itemize]{leftmargin=*}

\begin{document}

\title{Did NANOGrav see a signal from primordial black hole formation?}

\author{Ville Vaskonen}
\email{vvaskonen@ifae.es}
\affiliation{Institut de Fisica d’Altes Energies, The Barcelona Institute of Science and Technology, Campus UAB, 08193 Bellaterra, Barcelona, Spain}

\author{Hardi Veerm\"ae}
\email{hardi.veermae@cern.ch}
\affiliation{National Institute of Chemical Physics and Biophysics, R\"{a}vala 10, 10143 Tallinn, Estonia}

\begin{abstract}
We show that the recent NANOGrav result can be interpreted as a stochastic gravitational wave signal associated to formation of primordial black holes from high-amplitude curvature perturbations. The indicated amplitude and power of the gravitational wave spectrum agrees well with formation of primordial seeds for supermassive black holes.
\end{abstract}

\maketitle

\noindent\textbf{Introduction} -- Strong evidence for a stochastic common-spectrum process, that can be interpreted as a stochastic gravitational wave (GW) signal, was found in the recent analysis of 12.5-year pulsar timing array (PTA) data collected by the North American Nanohertz Observatory for Gravitational Waves (NANOGrav)~\cite{Arzoumanian:2020vkk}. NANOGrav observes a narrow range of frequencies around $f = 5.5\,{\rm nHz}$. The potential GW signal can be fitted by a power-law $\Omega_{\rm GW} \propto f^\zeta$ with amplitude $\Omega_{\rm GW}(f = 5.5\,{\rm nHz})\in(3\times 10^{-10},2\times 10^{-9})$ and exponent $\zeta\in(-1.5,0.5)$ at $1\sigma$ confidence level, and with a small positive correlation between the amplitude and the exponent.

A possible source for a stochastic GW background at such frequencies is supermassive black hole (SMBH) binary inspirals~\cite{Sesana:2004sp}, which give $\Omega_{\rm GW}\propto f^{2/3}$. Their merger rate and therefore the resulting amplitude of the GW signal has, however, large uncertainties. Alternatively, instead of being astrophysical, a strong stochastic GW background at nanoHerz frequencies can originate from cosmological sources. For example, the NANOGrav result has been recently interpreted as a signal from cosmic strings~\cite{Ellis:2020ena,Blasi:2020mfx}. 

PTA experiments are sensitive to parts of the secondary GW background associated with the production of planetary mass or heavier primordial black holes (PBHs) from large curvature perturbations.\footnote{We note that PTAs can not probe the mass window below $10^{-10}\Msun$ in which PBHs may constitute all dark matter (DM), as the formation of these PBHs corresponds to much higher frequencies.} They may therefore probe two open problems:  First, it is so far unknown whether the black hole (BH) binaries observed by the LIGO/Virgo collaboration~\cite{LIGOScientific:2018mvr,Abbott:2020khf,Abbott:2020tfl} are of astrophysical or primordial origin. Although scenarios in which PBHs in the solar mass range comprise all of DM are heavily constrained~\cite{Carr:2020gox,Vaskonen:2019jpv,DeLuca:2020jug}, they might still account for the LIGO/Virgo BH mergers when they make up about 0.1\%\,--\,10\% of the DM density~\cite{Bird:2016dcv,Clesse:2016vqa,Sasaki:2016jop,Raidal:2017mfl,Ali-Haimoud:2017rtz,Raidal:2018bbj,Vaskonen:2019jpv,DeLuca:2020jug}. Second, PBHs heavier than $10^3 \Msun$ can provide seeds for SMBHs~\cite{Duechting:2004dk,Kawasaki:2012kn,Carr:2018rid} and act as generators for cosmic structures~\cite{Carr:2018rid}. In particular, the origin of SMBHs has been a long-standing problem in astrophysics as, although their existence at the center of most galaxies has been well established~\cite{Kormendy:1995er,Magorrian:1997hw,Richstone:1998ky}, their astrophysical production seems to require super-Eddington accretion~\cite{2016MNRAS.458.3047P} or direct collapse into intermediate mass BHs~\cite{2017MNRAS.468.3935H}. In the PBH scenario, even a small abundance of heavier than $10^3\Msun$ PBHs can provide the seeds for SMBHs.

In this Letter we interpret the NANOGrav result as a stochastic GW background associated to PBH formation from high-amplitude peaks in the primordial curvature power spectrum. We consider two different well motivated shapes for a peak in the curvature power spectrum and, assuming the standard radiation dominated expansion history, we calculate the secondary GW spectrum and the corresponding PBH abundance and mass function.

\vspace{4pt}\noindent\textbf{Peaks in the curvature power spectrum} -- In order to perform our analysis in a model independent fashion, we consider two different shapes for the peak in the curvature power spectrum. 

First, typical peaks generated in single field inflation~\cite{Ballesteros:2017fsr,Byrnes:2018txb} can be approximated by a broken power-law
\be \label{eq:Pk_PL}
    \mathcal{P}_{\rm PL}(k) = A \,\frac{\alpha+\beta}{\beta (k/k_*)^{-\alpha} + \alpha (k/k_*)^{\beta}} \,,
\ee
where $\alpha, \beta>0$ describe respectively the growth and decay of the spectrum around the peak. In single field models, where a peak is generated via a quasi-inflection point, one typically has $\alpha \lesssim 4$~\cite{Byrnes:2018txb,Carrilho:2019oqg}. Additionally, it follows that $\beta \gtrsim 0.5$, if the curvature power spectrum between the end of inflation and the peak obeys a power law.\footnote{This follows from $k_{\rm end}<10^{23}\,\Mpc^{-1}$ and $\mathcal{P}_{\rm PL}(k_{\rm end})<H_{\rm inf}^2/(8\pi \Mpl^2)<2.5\times 10^{-11}$~\cite{Akrami:2018odb} and a peak with $A < 0.05$ at $k_{*} > 10^4\,\Mpc^{-1}$.} As a benchmark case we take in the following $\alpha=4$ and $\beta=0.5$.

Second, we consider a log-normal peak with an exponential UV cut-off,
\be \label{eq:Pk_LN}
\mathcal{P}_{\rm LN}(k)\!=\!A \exp\left[\beta\left(1\!-\!\frac{k}{k_*}\!+\!\ln\!\left(\frac{k}{k_*}\right)\right)\!-\!\alpha\ln^2\!\left(\frac{k}{k_*}\right)\right],
\ee
where $\alpha, \beta>0$. For example, with $\alpha = 0.17$ and $\beta = 0.62$ this shape fits well the peak obtained in two field inflation considered in Ref.~\cite{Braglia:2020eai}, and we therefore use these values as a benchmark case.

From a theoretical perspective, peaked primordial power spectra required for producing above planetary mass PBH are likely not realised in the simplest inflationary models as such models tend to generate too low spectral index $n_s$ for the CMB~\cite{Kannike:2017bxn,Ballesteros:2017fsr,Braglia:2020eai}. This is because of the short period of less than 20 $e$-folds between the peak and the CMB scales, which in a wide range of single field inflation models would produce $n_s = 1 - \mathcal{O}(1)/\Delta N \lesssim 0.95$~\cite{Kannike:2017bxn} in strong tension with the CMB observations, $n_s = 0.966(4)$~\cite{Akrami:2018odb}. However, such issues can be avoided by sharp features in the scalar field evolution, e.g. sudden turns in two field space. 

\begin{figure}
\centering
\includegraphics[width=0.95\columnwidth]{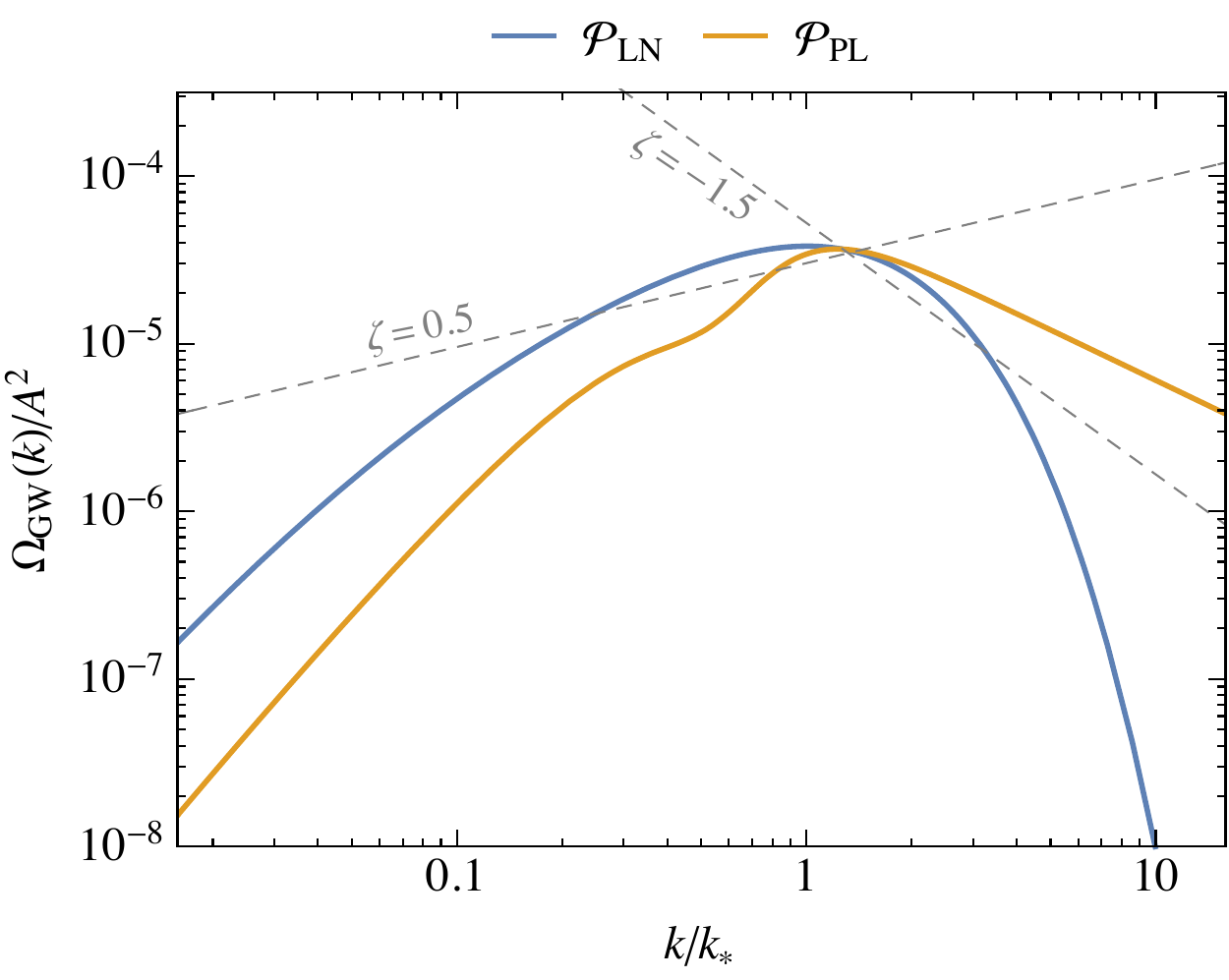}
\caption{The SIGW spectrum for the curvature power spectra given in \eqref{eq:Pk_PL} and \eqref{eq:Pk_LN}. The dashed lines indicate the $1\sigma$ NANOGrav region $\zeta\in(-1.5,0.5)$.}
\label{fig:OmegaGW}
\end{figure}

\vspace{4pt}\noindent\textbf{Scalar induced GWs} -- Curvature perturbations induce formation of GWs at second order from mode coupling~\cite{Matarrese:1993zf,Matarrese:1997ay,Nakamura:2004rm,Ananda:2006af,Baumann:2007zm}. Recently these scalar-induced GWs (SIGWs) have been extensively studied, and the prospects for observing them have been considered in Refs.~\cite{Saito:2008jc,Assadullahi:2009jc,2011PhRvD..83h3521B,Inomata:2018epa,Byrnes:2018txb,Clesse:2018ogk,Wang:2019kaf,Chen:2019xse}. 

During radiation domination GWs decouple from scalar perturbations soon after horizon crossing and their abundance reaches a constant value. In Ref.~\cite{Inomata:2019yww} (see also Refs.~\cite{DeLuca:2019ufz,Yuan:2019fwv}) it was shown that the observable SIGW background produced during radiation domination is gauge independent. The SIGW spectrum today is given by~\cite{Kohri:2018awv,Espinosa:2018eve,Inomata:2019yww}
\begin{widetext}
\be
    \Omega_{\rm GW}(k) = 0.387\, \Omega_{\rm R} \left(\frac{g_{*,s}^{4}g_{*}^{-3}}{106.75}\right)^{-\frac{1}{3}}
    \frac{1}{6} \int_{-1}^1 \td x \int_1^\infty \td y \, \mathcal{P}\left(\frac{y-x}{2}k\right) \mathcal{P}\left(\frac{x+y}{2}k\right) F(x,y) \,,
\ee
where $\Omega_{\rm R} = 5.38\times10^{-5}$ is the radiation abundance~\cite{Aghanim:2018eyx}, the effective numbers of degrees of freedom, $g_{*,s}$ and $g_*$, are evaluated at the moment when the constant abundance is reached, roughly coinciding with the horizon crossing moment, and
\bea
    F(x,y) = &\frac{288(x^2+y^2-6)^2(x^2-1)^2(y^2-1)^2}{(x-y)^8(x+y)^8} \\ &\hspace{10mm}\times\left[\left(x^2-y^2+\frac{x^2+y^2-6}{2}\log\left|\frac{y^2-3}{x^2-3}\right|\right)^2 + \frac{\pi^2}{4}(x^2+y^2-6)^2\theta(y-\sqrt{3}) \right] \,.
\eea
\end{widetext}

Examples of different SIGW spectra are shown in Fig.~\ref{fig:OmegaGW}. The amplitude of the spectrum depends very weakly on $k_*$; only through the effective number of degrees of freedom. In Fig.~\ref{fig:OmegaGW} we used $k_*=3.6\times 10^6\,\Mpc^{-1}$ which corresponds to the temperature $T \approx 0.2\,\GeV$. The position of the peak of the spectrum is determined by $k_*$ and its amplitude inherits its scaling from the curvature power spectrum peak as $\Omega_{\rm GW} \propto A^2$. For a power-law curvature power spectrum $\mathcal{P}\propto k^{\zeta/2}$, the SIGW spectrum behaves as $\Omega_{\rm GW}\propto k^\zeta$.

\vspace{4pt}\noindent\textbf{PBH formation} -- Consider a fluctuation with density contrast $\delta_{m}$ at comoving scale $k$. In radiation dominated Universe an overdensity for which $\delta_{m}$ is larger than threshold value $\delta_c$ part of the horizon mass,
\be \label{eq:M_k}
    M_k 
    \approx 1.4 \times 10^{13}\Msun \left(\frac{k}{\Mpc^{-1}} \right)^{-2}\, \left(\frac{g_{*,s}^{4}g_{*}^{-3}}{106.75}\right)^{-1/6} \,,
\ee  
collapses to BH almost immediately when the scale $k$ reenters horizon~\cite{Carr:1975qj}. The masses of the produced PBHs follow the critical scaling~\cite{Choptuik:1992jv,Niemeyer:1997mt,Niemeyer:1999ak}
\be\label{eq:Mcrit}
    M = \kappa M_k \left(\delta_{m}  - \delta_c \right)^{\gamma} \,,
\ee
where $\gamma = 0.36$ corresponds to the universal critical exponent during radiation domination~\cite{Choptuik:1992jv,Evans:1994pj}, and we must account for the nonlinear relation $\delta_{m} = \delta_{\zeta} - \frac{3}{8}\delta_{\zeta}^2$ between the density contrast and the curvature perturbation $\delta_{\zeta}$~\cite{Young:2019yug,DeLuca:2019qsy,Kawasaki:2019mbl}. The critical threshold $\delta_c$ for PBH formation and the $\kappa$ parameter depend on the procedure used to smooth the primordial perturbations~\cite{Young:2019osy,Young:2020xmk,Gow:2020bzo} as well as on the shape of individual peaks~\cite{Musco:2018rwt,Young:2019yug,Escriva:2020tak}. Nonsphericity was found to have an insignificant effect on the threshold for large perturbations~\cite{Yoo:2020lmg}.

The fraction of the total energy density $\beta_k(M)\td \ln M$ that collapses into BHs of mass $M$ can be estimated using the Press-Schechter formalism~\cite{Press:1973iz,Bond:1990iw,Carr:1975qj,Gow:2020bzo}. It gives
\bea \label{eq:betak}
    \beta_k(M)
&    = \int_{\delta_c}^\infty \td\delta\, \frac{M}{M_k} \, P_k(\delta) \delta_D\left[ \ln\frac{M}{M(\delta)} \right]  \\
&    = \frac{2\kappa}{\gamma} \frac{q^{1+1/\gamma} P_k(\delta_{\zeta}(M)) }{1 - \frac{3}{4}\delta_{\zeta}(M)} 
    \,,
\eea
where $\delta_D$ denotes the Dirac delta function and $\delta_{\zeta}(M) = 4\left[1 - \sqrt{1 - \frac{3}{2}\left(\delta_c + q^{1/\gamma}\right)}\right]/3$ is the inversion of Eq.~\eqref{eq:Mcrit}, $q \equiv M/(\kappa M_k)$. We assume a Gaussian distribution for the curvature perturbations, 
\be\label{eq:Pk}
    P_k(\delta) = \frac{1}{\sqrt{2\pi} \sigma_k} \exp\left( -\frac{\delta^2}{2\sigma_k^2} \right) \,,
\ee 
where the variance $\sigma_k^2$ is
\be \label{eq:sigmak}
    \sigma_k^2 = \!\left(\frac{4}{9}\right)^2\!\!\int_0^\infty \frac{\td k'}{k'}\!\left(\frac{k'}{k}\right)^{\!4}\! W^2(k'/k)\, T^2(k'/k) \mathcal{P}(k') .
\ee
Following Ref.~\cite{Young:2019yug}, we will use a real-space top-hat window function $W(k'/k)$ and account for the damping of sub-horizon fluctuations with the linear transfer function $T(k'/k)$.\footnote{This implies that the evolution of sub-horizon fluctuations is linear, which might not be the case for large fluctuations.}

At the present day, the PBH mass function normalised to the total PBH abundance, $\int \td \ln M \psi(M) = \Omega_{\rm PBH}$, is
\bea \label{eq:psiRD}
     \psi(M) 
&     = \int \td\ln k \, \beta_k(M) \frac{\rho_{\gamma}(T_k)}{\rho_c} \frac{s(T_0)}{s(T_k)} \, \\    
&     \simeq \frac{4 \times 10^{-12}}{\gamma} \!\frac{M}{\Msun} \int \!\frac{\td k \,k^2}{\Mpc^{-3}}  \frac{q^{1/\gamma}P_k(\delta_{\zeta}(M))}{1 - \frac{3}{4}\delta_{\zeta}(M)}  \,,
\eea
where $\rho_{\gamma}(T)$ and $s(T)$ denote the energy and entropy densities of radiation at temperature $T$, and $\rho_c$ is the critical energy density of the Universe. By numerical fits we find that the PBH mass function for the curvature power spectra \eqref{eq:Pk_PL} and \eqref{eq:Pk_LN} is roughly of the form
\be \label{eq:mf}
    \psi(M)\propto M^{1+1/\gamma} e^{-c_1 (M/\langle M_{\rm PBH} \rangle )^{c_2}} \,,
\ee
where $c_1$ is fixed by demanding that $\langle M_{\rm PBH} \rangle$ is the average PBH mass, and $c_2 \simeq 1$ depends mildly on the amplitude of the peak. The low mass tail of the mass function is dominated by PBHs forming close to the threshold and is thus determined by the details of the critical collapse~\cite{Niemeyer:1997mt}. The heavier tail gets exponentially suppressed as density perturbations capable of producing heavier PBHs become exponentially more unlikely. The abundance of PBHs and their mean mass are
\bea \label{eq:fits}
    \Omega_{\rm PBH} &\simeq c_\Omega A^{c'_\Omega}\, e^{-c_A/A}\, k_*/\Mpc^{-1} \,,
\\ 
    \left \langle M_{\rm PBH} \right\rangle &\simeq c_M A^{c'_M} M_{k_*} \,,
\eea
where for $\delta_c = 0.55$ and $\kappa=4$ we find $c_\Omega \approx 100$, $c'_\Omega \approx 1.3$, $c_A \approx 0.3$, $c_M \approx 10$ and $c'_M \approx 1/3$. In the following we show the PBH abundance relative to the observed DM abundance, $f_{\rm PBH}\equiv \Omega_{\rm PBH}/\Omega_{\rm DM}$, where $\Omega_{\rm DM}=0.26$~\cite{Aghanim:2018eyx}.

\begin{figure}
\centering
\includegraphics[width=\columnwidth]{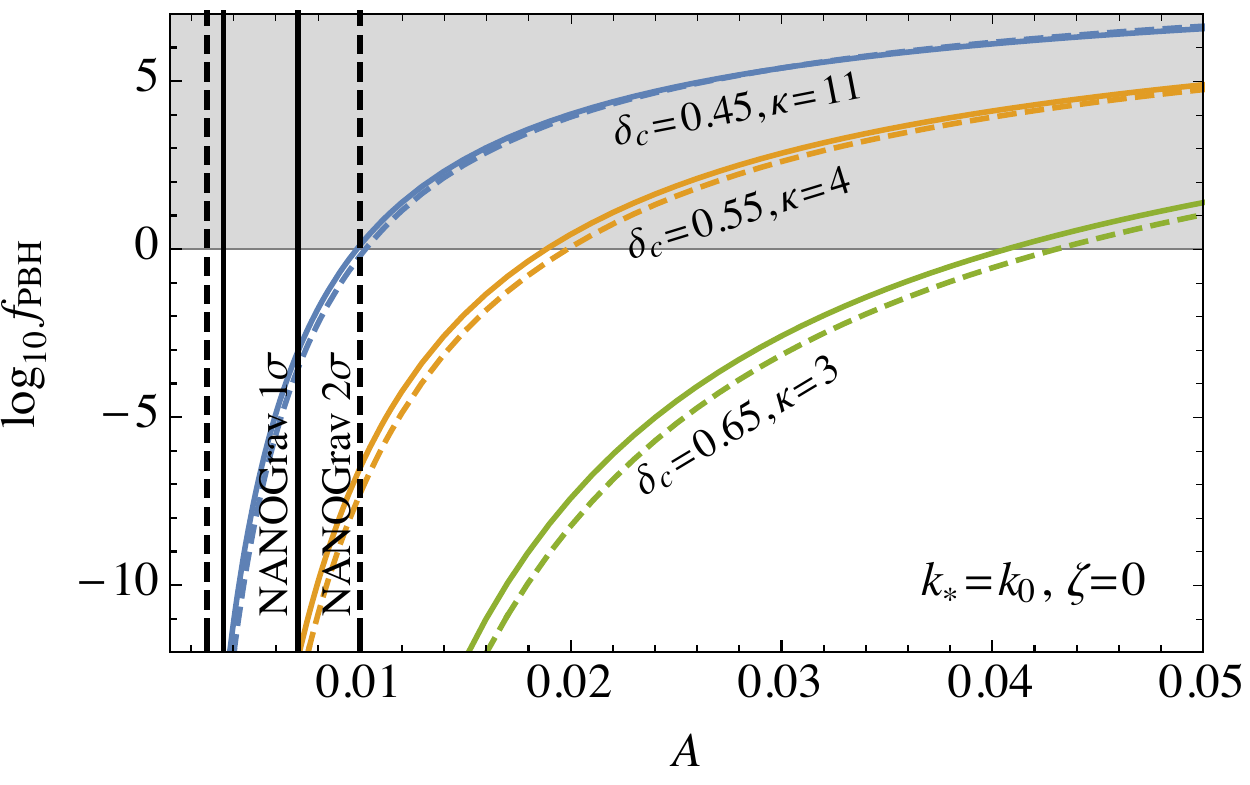}
\caption{Abundance of PBHs as a function of the amplitude of a log-normal (solid) and broken power-law (dashed) peaks in the curvature power spectrum at $k_* = k_0$. The black solid and dashed contours indicate the $1\sigma$ and $2\sigma$ NANOGrav ranges for $\zeta=0$.}
\label{fig:deltac}
\end{figure}

The PBH abundance is produced from the tail of a Gaussian distribution and thus even $\mathcal{O}(10\%)$ changes to the threshold can correspond to order of magnitude differences in the PBH abundance. This is illustrated in Fig.~\ref{fig:deltac} where we show the PBH abundance for $(\delta_c,\kappa) = (0.45,11),\,(0.55,4),\,(0.65,3)$ corresponding to the critical collapse of differently shaped peaks~\cite{Young:2019yug}. We note that the theoretically allowed range can be slightly wider, $\delta_c \in (0.41,2/3)$~\cite{Musco:2018rwt}. In the following numerical estimates we use $\delta_c = 0.55$ and $\kappa=4$.

An alternative prescription based on peaks theory replaces the transfer function with a hard cut-off of the top-hat window function~\cite{Gow:2020bzo}. In this case it was shown that different window functions combined with a careful matching of the critical collapse parameters does affect the required height of the peak in the curvature power spectrum for a given PBH abundance by $\mathcal{O}(10\%)$. However, as the position of the hard cut-off can introduce additional errors, we rely on the prescription given in Ref.~\cite{Young:2019yug}, which is more directly related numerical results on critical collapse, although the use of the linear transfer function may not be completely justified in this case. In all, the theoretical uncertainties related to the choice in the window function can be effectively absorbed by the uncertainties in the critical collapse parameters resulting in an even greater variation in $f_{\rm PBH}$ than shown in Fig.~\ref{fig:deltac}.

Our PBH abundance estimates may additionally be affected by variations in the shape of the peak in the curvature power spectrum in specific inflationary models, by non-Gaussianities~\cite{Franciolini:2018vbk,Atal:2018neu,Atal:2019cdz,DeLuca:2019qsy,Kehagias:2019eil,Yoo:2019pma}, or by changes to the equation of state of the thermal bath, e.g. during the QCD phase transition~\cite{Jedamzik:1996mr,Byrnes:2018clq}.

\vspace{4pt}\noindent\textbf{Results} -- The first five bins of the NANOGrav analysis, for which a power-law fit is provided in~\cite{Arzoumanian:2020vkk}, are in the narrow frequency range $f/{\rm Hz}\in(2.5\times 10^{-9},1.2\times 10^{-8})$. Therefore, we expand the predicted spectrum around $k_0 = 2\pi\times 5.5\,{\rm nHz} = 3.6\times 10^6\,\Mpc^{-1}$ as
\be \label{eq:expansion}
    \Omega_{\rm GW} \simeq \Omega_{\rm GW,0} \left(k/k_0\right)^{\zeta} \,,
\ee
and compare the experimental ranges for the parameters $\Omega_{\rm GW,0}$ and $\zeta \equiv \td\ln\Omega_{\rm GW}(k_0)/\td\ln k$ with the theoretical predictions of the SIGW for a given primordial curvature spectra.

The SIGW spectrum has a flat region around $k\sim k_*$ and can thus provide a good fit for the shape of the NANOGrav signal. From Fig.~\ref{fig:OmegaGW} we see that around the peak $\Omega_{\rm GW} \simeq 4\times10^{-5} A^2$, while Eq.~\eqref{eq:fits} with $k_*=k_0$ shows that PBHs will comprise a significant fraction of DM if 
$A \simeq 0.02$. This implies $\Omega_{\rm GW} \simeq 2 \times 10^{-8}$ above the NANOGrav $2\sigma$ region. Moreover, Eq.~\eqref{eq:fits} indicates that the corresponding mean PBH mass is $\langle M_{\rm PBH}\rangle \simeq 3\Msun$.

\begin{figure}
\centering
\includegraphics[width=\columnwidth]{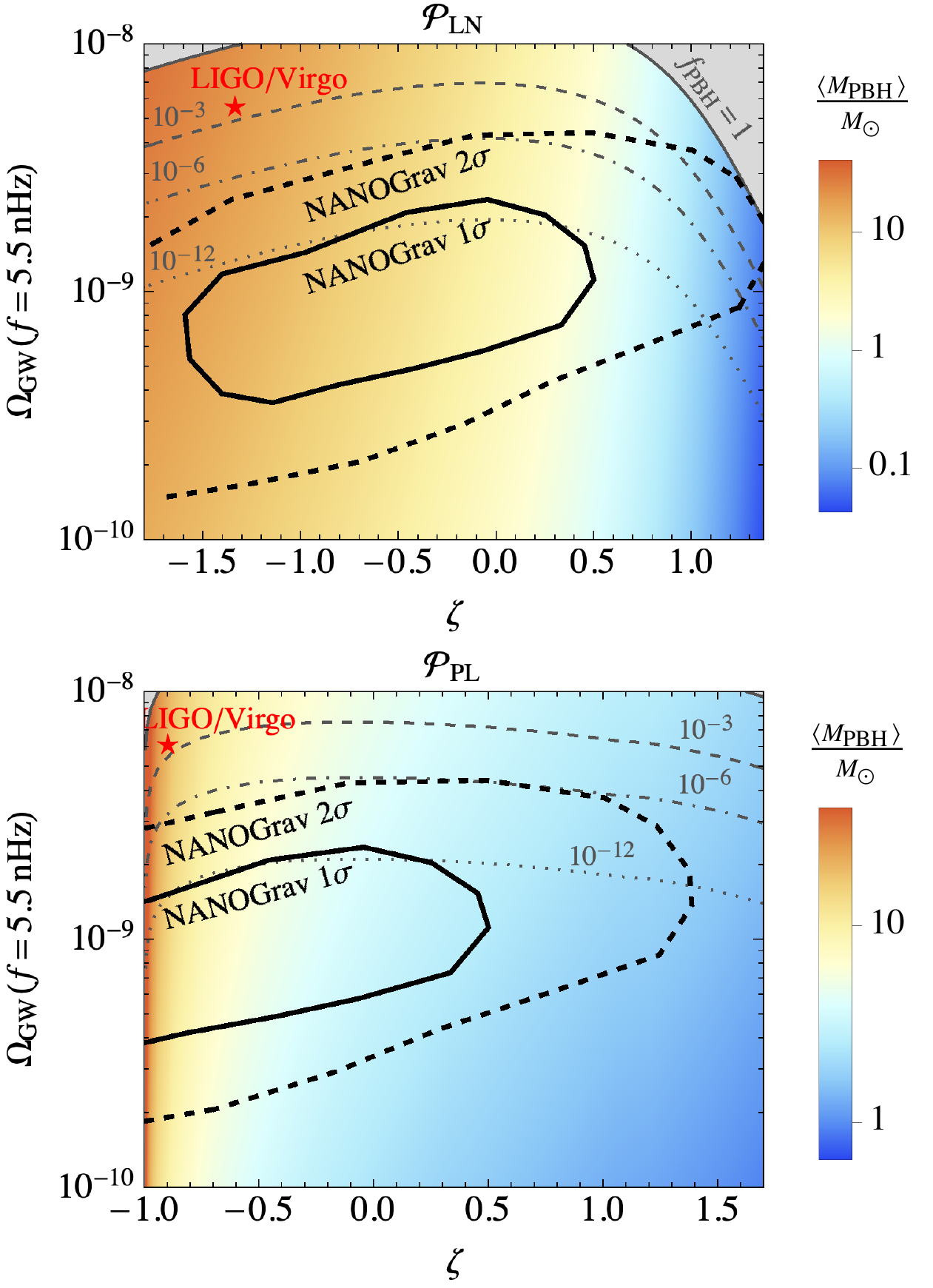}
\caption{The thick black solid and dashed contours show the $1\sigma$ and $2\sigma$ ranges for the power $\zeta$ and amplitude $\Omega_{\rm GW}(f=5.5\,{\rm nHz})$ of the GW spectrum indicated by the NANOGrav results~\cite{Arzoumanian:2020vkk}. The thin solid, dashed and dot-dashed lines instead show the PBH abundance and the color coding shows the mean mass of the PBH mass spectrum for the curvature power spectra~\eqref{eq:Pk_PL} and~\eqref{eq:Pk_LN}. The gray regions are excluded by overproduction of PBHs and the red star indicates the PBH scenario for the LIGO/Virgo events.}
\label{fig:NG}
\end{figure}

This is illustrated in Fig.~\ref{fig:NG}, where we have fixed $k_*$ and $A$ such that the expansion~\eqref{eq:expansion} holds around $k=k_0$ which, for a given shape of the curvature power spectrum peak, fixes the PBH abundance and mass function. The thick contours show the $1\sigma$ and $2\sigma$ confidence level regions indicated by the NANOGrav result, obtained by a simple transformation~\cite{Ellis:2020ena} from the power-law fit to the five lowest frequency bins presented in~\cite{Arzoumanian:2020vkk}. We see that the $1\sigma$ region barely crosses the dotted gray curve indicting $f_{\rm PBH} = 10^{-12}$. The mean PBH mass is indicated by the color coding, which shows that for $\zeta> 0$ the PBH masses are mostly $\langle M_{\rm PBH}\rangle\lesssim 10\Msun$. For $\zeta<0$ heavier PBH will be formed as the peak in the curvature power spectrum lies at $k<k_0$.

\begin{figure}
\centering
\includegraphics[width=\columnwidth]{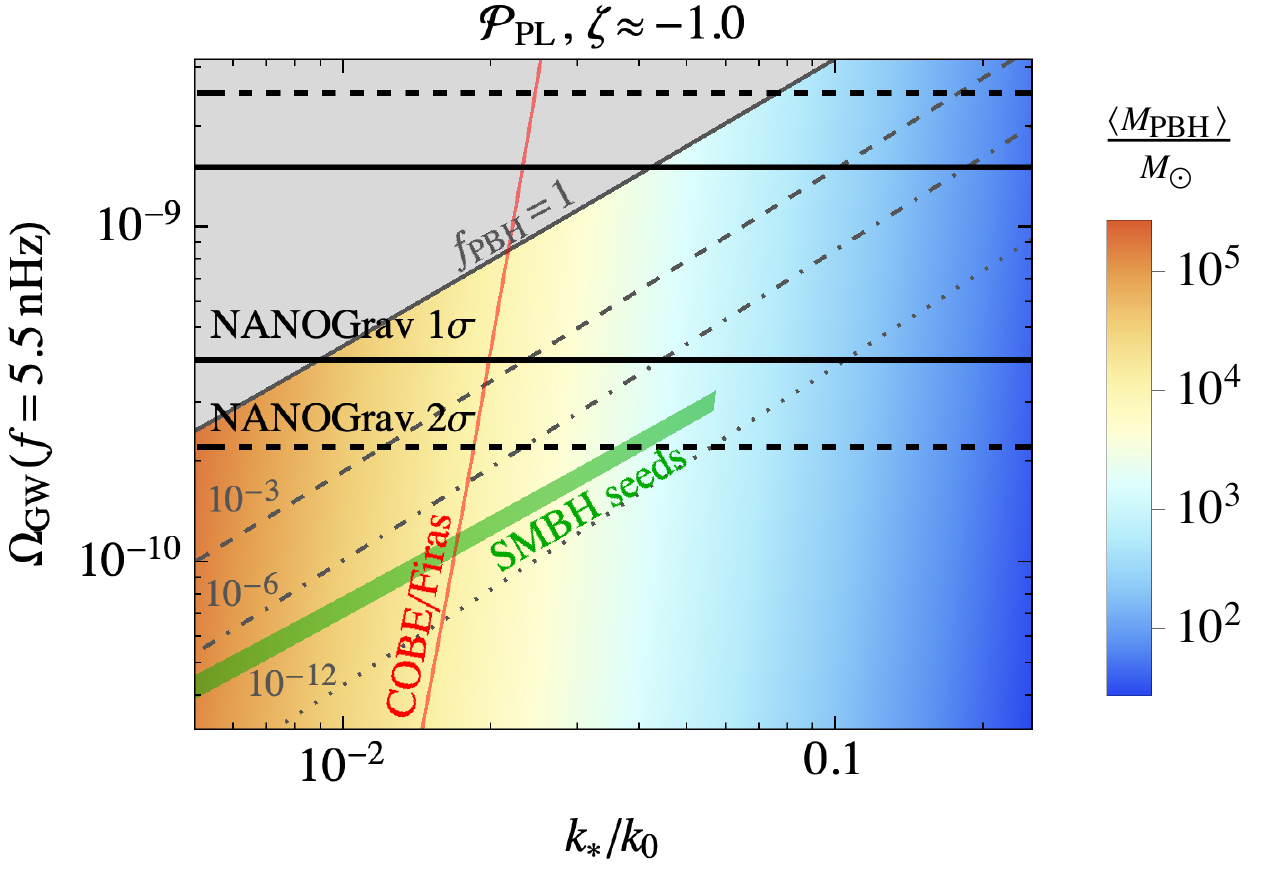}
\caption{The thick black solid and dashed lines indicate the $1\sigma$ and $2\sigma$ lower boundaries of the NANOGrav range for the amplitude $\Omega_{\rm GW}(f=5.5\,{\rm nHz})$ at $\zeta = -1.0$. The thin solid, dashed and dot-dashed lines instead show the PBH abundance and the color coding shows the mean mass of the PBH mass function, assuming a power-law peak in the curvature power spectrum~\eqref{eq:Pk_PL} with $\alpha=4$ and $\beta=0.5$. The gray region is excluded by overproduction of PBHs. The region left of the red line is excluded by the COBE/Firas results on the CMB $\mu$ distortions. The green band indicates the values compatible with SMBH seed formation.}
\label{fig:beta05}
\end{figure}

Because of the experimental uncertainties in the slope of the stochastic GW signal, it is possible that the peak of the SIGW spectrum lies away from the NANOGrav range, especially if the SIGW spectrum has relatively flat tails. For the log-normal benchmark curvature spectra the tails are too steep, and the slopes compatible with the NANOGrav range are near the peak of the spectrum. However, as can be seen from Fig.~\ref{fig:OmegaGW}, for the broken power-law benchmark case the slope of the high frequency tail, $\zeta = -2\beta = -1.0$, is within the $1\sigma$ region. 

Fig.~\ref{fig:beta05} shows the SIGW amplitude at the NANOGrav frequency $f_0$ for the broken power-law benchmark case as a function of $k_*<0.25k_0$. In this case, the NANOGraw signal is generated from the relatively flat high $k$ tail ($\zeta\approx -1.0$) of the power spectrum, while PBHs are dominantly produced by the peak at $k_*$. A sizable PBHs abundance is consistent with the $2\sigma$ NANOGrav region when $k_* \lesssim 0.1k_0$. Fig.~\ref{fig:beta05} also indicates the COBE/Firas bound from CMB $\mu$ distortions~\cite{Fixsen:1996nj,Chluba:2012we}, which excludes the region left from the red line. A milder slope of the low-$k$ tail of the curvature power spectrum peak, $\alpha<4$, would move the COBE/Firas bound towards higher $k_*$ and thus closer to the region consistent with the NANOGrav signal. We also remark that future PIXIE like experiments~\cite{Kogut:2011xw} may probe the SIGW interpretation of the NANOGrav signal.

As was outlined in the Introduction, there are two PBH scenarios with a particular phenomenological relevance:
\begin{itemize}
\setlength\itemsep{-0.5mm}
    \item \emph{Primordial origin for the LIGO/Virgo BH mergers} requires a distribution of PBH with~\cite{Vaskonen:2019jpv}
    \be
        f_{\rm PBH} \sim 0.01\,, \quad
        \langle M_{\rm PBH} \rangle \approx 20 \Msun \,,
    \ee
    and a narrow width of the PBH mass function. As indicated by the red star in Fig.~\ref{fig:NG}, we find that scenarios with broken power-law peaks would require a SIGW background that is too strong to be compatible with the NANOGrav signal. Changing the slope of the primordial curvature power spectrum will not relieve this tension. Since the allowed values of $A$ decrease with $\zeta$, picking a bigger value for $\beta$ would not put the scale $k_{*}$ within the $2\sigma$ region. Nevertheless, the nontrivial modifications of the shape of the spectrum and uncertainties associated with the critical collapse, illustrated in Fig.~\ref{fig:deltac}, may change the theoretical abundance estimate by several orders of magnitude and thus relieve the tension within common scenarios for LIGO/Virgo PBH and the NANOGrav signal.
    
    \item For \emph{primordial origin for SMBH seeds} we assume a mass range of $M_{\rm PBH} \in (10^3, 10^{6}) \Msun$~\cite{Duechting:2004dk}. To roughly estimate the required seed abundance, we assume that the SMBHs comprise about 0.025\% of the stellar mass in their host galaxies~\cite{Reines_2015}, while stars make up a fraction of about 1\% of the Universes matter content~\cite{Fukugita:1997bi}. This implies that the total SMBH density is about a factor of $10^{6}$ smaller than the DM density. Using $10^{7}\Msun$ as a representative value for the SMBH mass, we find that their primordial seeds can be characterized by
    \be
        f_{\rm PBH} \sim 10^{-13} \langle M_{\rm PBH} \rangle/\Msun, \quad
        \langle M_{\rm PBH}\rangle > 10^3 \Msun.
    \ee
    Fig.~\ref{fig:beta05} shows that production of a sufficiently large abundance of primordial SMBH seeds can be consistent with the NANOGrav signal and the $\mu$-distortion constraints.
\end{itemize}

\vspace{4pt}\noindent\textbf{Conclusions} -- We showed that the NANOGrav result can be interpreted as a signal from PBH formation from peaks in the curvature power spectrum. We found that the secondary GW backgrounds consistent with NANOGrav will, in general, correspond to the production of a negligible amount of PBH DM and is thus in tension with the PBH scenario for LIGO/Virgo merger events. However, this tension might be relieved when accounting for theoretical uncertainties in PBH formation. The NANOGrav signal agrees well with scenarios in which PBHs provide the seeds of supermassive black holes.

Although our estimates provide a viable proof on concept for PBH scenarios related to NANOGrav, stronger and more definite conclusions can be drawn only by reducing the theoretical uncertainties related to PBH formation and require dedicated computations based on specific inflationary scenarios. However, as the formation of any amount of PBHs is allowed only within a narrow range of amplitudes of the power spectrum, it is intriguing that the NANOGrav signal is consistent with producing even a small PBH abundance.

\vspace{4pt}\noindent\emph{Acknowledgments} -- We thank Chris Byrnes, Gert H\"utsi, Marek Lewicki and Sam Young for helpful discussions. This work was supported by the European Regional Development Fund through the CoE program grant TK133, the Mobilitas Pluss grants MOBTP135, MOBTT5 and the Estonian Research Council grant PRG803. The work of VV is supported by Juan de la Cierva fellowship from Spanish State Research Agency.

\vspace{4pt}\noindent\emph{Note added:} -- Soon after the first version of this work, Ref.~\cite{Kohri:2020qqd} claimed that the NANOGrav signal may be consistent with the LIGO/Virgo PBH scenario attributing the discrepancy between our conclusions to the choice of window function used in Eq.~\eqref{eq:sigmak}. We improved our PBH abundance estimate following~\cite{Young:2019osy} and found that our conclusions remain in tact. The differences between our conclusions and Ref.~\cite{Kohri:2020qqd} can be resolved when we omit the nonlinear relation between the density contrast and curvature perturbations. Additionally, Ref.~\cite{DeLuca:2020agl} appeared pointing out a potential scenario for light PBH DM consistent with NANOGrav, which was not considered here.

\bibliography{PBH}
\end{document}